\documentstyle[12pt]{article}

\def\noi{\noindent}

\def\nqq{\hspace{-2em}}

\def\barr{\left(\begin{array}}
\def\earr{\end{array}\right)}
\def\beq{\begin{equation}}
\def\eeq{\end{equation}}
\def\ber{\begin{eqnarray} &&\nqq}
\def\eer{\end{eqnarray}}
\def\eern{\nonumber \end{eqnarray}}
\def\nn{\nonumber\\ &&\nqq}
\def\mm{\\ &&\nqq}

\catcode`\@=11 \@addtoreset{equation}{section}\catcode`\@=12

\newcommand{\const}{\mathop{\rm const}\nolimits}

\newcommand{\ch}{\mathop{\rm ch}\nolimits}
\newcommand{\Ric}{\mathop{\rm Ric}\nolimits}
\newcommand{\btd}{\bigtriangledown}
\newcommand{\btu}{\bigtriangleup}
\newcommand{\eps}{\varepsilon}
\newcommand{\til}{\tilde}
\newcommand{\e}[1]{\mathop{\rm e}\nolimits^{#1}}
\newcommand{\ints}{\int\limits}
\newcommand{\half}{\frac{1}{2}}

\newcommand{\p}{\partial}

\newcommand{\ds}{\displaystyle}

\newcommand{\fnm}{\footnotemark}
\newcommand{\fnt}{\footnotetext}

\begin{document}


\begin{center}
\large\bf
MULTIDIMENSIONAL COSMOLOGY FOR INTERSECTING P-BRANES WITH STATIC \\
INTERNAL SPACES\\[25pt]

\normalsize\bf
M.A. Grebeniuk\fnm[1]\fnt[1]{mag@gravi.phys.msu.su} \\[5pt]

\it
Moscow State University, Physical Faculty,
Department of Theoretical \\ Physics, Moscow 117234, Russia \\[10pt]

\bf
V.D. Ivashchuk\fnm[2]\fnt[2]{ivas@rgs.phys.msu.su}
and V.N. Melnikov\fnm[3]\fnt[3]{melnikov@rgs.phys.msu.su} \\[5pt]

\it
Centre for Gravitation and Fundamental Metrology,
VNIIMS \\ 3-1, M.Ulyanovoy Str., Moscow 117313, Russia
\end{center}

\vspace{25pt}

\small\noi
Multidimensional cosmological model with static internal spaces
describing the evolution of an Einstein space of non-zero curvature and
$n$ internal spaces is considered. The action contains several dilatonic
scalar fields $\varphi^I$ and antisymmetric forms $A^I$ and $\Lambda$-term.
When forms are chosen to be proportional to volume forms of $p$-brane
submanifolds of internal space manifold, the Toda-like Lagrange
representation arises. Exact solutions for the model are obtained, when
scale factors of internal spaces are constant. It is shown that they are
de Sitter or anti-de Sitter. Behaviour of cosmological constant and its
generation by p-branes is demonstrated.

\normalsize

\section{Introduction}

Here we consider a multidimensional gravitational model governed by the
action containing several dilatonic scalar fields and antisymmetric forms
\cite{GrIM,BGIM,IM} and $\Lambda$-term with the aim of studying the solutions
with static internal spaces. They are important if we want to have effective
gravitational constant be really constant.

We study a cosmological sector of the model from \cite{IMO} (see also
\cite{IMC}). We recall that the model from \cite{IMO} incorporates generalized
non-composite intersecting $p$-brane solutions. Using the $\sigma$-model
representation from \cite{IMO,IMC} we reduce equations of motion to the
pseudo-Euclidean Toda-like Lagrange system \cite{IM3}--\cite{GIM} with the
zero-energy constraint.

Here we consider the case of constant vectors \cite{BIMZ,ZHUK1,ZHUK} in the
Toda potential and obtain exact solutions for the system with $p$-branes.
In this case we deal with the set of algebraic equations and only one
differential equation for $M_0$ space. We investagate two possible cases,
when all $p$-branes "live" on external manifold, and when $p$-brane
world-sheet do not cover $M_0$ space.

In the first case using the synchronous-time parametrization we get
the general exact solution for the initial cosmological metric. To
get this solution we also use the so-called "fine-tuning" of the
cosmological constant and curvatures of internal spaces. The second
case turns out to be off interest due to independence of the solution
on any $p$-brane configurations. In this case we have an ordinary
multidimensional cosmological solution with $\Lambda$-term \cite{BIMZ,ZHUK1}.
We note that the solution with de Sitter and anti-de Sitter spaces are of
interest now due to recent papers on M-theory on AdS/Superconformal theory
duality \cite{Mald,Witt,Poly}.

\section{The model}

Here like in \cite{IMO} we consider the model governed by the action
\ber
\label{2.1}
S=\frac{1}{2\kappa^2}
\int_{M}d^Dz\sqrt{|g|}\biggl\{R[g]-2\Lambda-
\sum_{I\in\Omega}\Bigl[g^{MN}\partial_{M}\varphi^I
\partial_{N}\varphi^I+\frac{1}{n_I!}\exp(2\lambda_{JI}\varphi^J)
(F^I)^2_g\Bigr]\biggr\}, \nn
\eer
where $g=g_{MN}dz^{M}\otimes dz^N$ is a metric, $\varphi^I$
is a dilatonic scalar field,
\beq
\label{2.2}
F^I=dA^I=\frac{1}{n_I!}F^I_{M_1\ldots M_{n_I}}
dz^{M_1}\wedge\ldots\wedge dz^{M_{n_I}}
\eeq
is a $n_I$-form ($n_I\ge2$) on $D$-dimensional manifold $M$,
$\Lambda$ is the cosmological constant and $\lambda_{JI}\in{\bf R}$,
$I,J \in \Omega$. In (\ref{2.1}) we denote $|g|=|\det(g_{MN})|$,
\beq
\label{2.3}
(F^I)^2_g=F^I_{M_1\ldots M_{n_I}}F^I_{N_1\ldots N_{n_I}}
g^{M_1 N_1}\ldots g^{M_{n_I}N_{n_I}},
\eeq
$I\in\Omega$, Here $\Omega$  is a non-empty finite set.

Equations of motion corresponding to (\ref{2.1}) have the following
form
\ber
\label{2.4}
R_{MN}-\frac{1}{2}g_{MN}R=T_{MN}-\Lambda g_{MN}, \nn
T_{MN}=\sum_{I\in\Omega}\left[T_{MN}^{\varphi^I}+
\exp(2\lambda_{JI}\varphi^J)T_{MN}^{F^I}\right], \mm
\label{2.5}
{\btu}[g]\varphi^J-\sum_{I\in\Omega}\frac{\lambda_{JI}}{n_I!}
\exp\left(2\lambda_{KI}\varphi^K\right)(F^I)^2_g=0, \nn
{\btd}_{M_1}[g]\left(\exp(2\lambda_{KI}\varphi^K)
F^{I,M_1\ldots M_{n_I}}\right)=0,
\eer
$I,J \in \Omega$ and
\ber
\label{2.6}
T_{MN}^{\varphi^I}=\p_{M}\varphi^I\p_{N}\varphi^I-
\frac{1}{2}g_{MN}\p_{P}\varphi^I\p^{P}\varphi^I, \mm
\label{2.7}
T_{MN}^{F^I}=\frac{1}{n_{I}!}\biggl[-\frac{1}{2}g_{MN}(F^{I})^{2}_{g}+
n_{I}F^{I}_{MM_2\ldots M_{n_I}}F_{N}^{I,M_2\ldots M_{n_I}}\biggr].
\eer
In (\ref{2.5}) ${\btu}[g]$ and ${\btd}[g]$ are the Laplace-Beltrami and
covariant derivative operators respectively corresponding to $g$.

We consider the manifold
\beq
M={\bf R}\times M_{0}\times\ldots\times M_{n},
\eeq
with the metric
\beq
\label{2.8}
g=w\e{2{\gamma}(u)}du\otimes du+\sum_{i=0}^n\e{2\phi^i(u)}g^i,
\eeq
where $w=\pm1$, $u$ is a time variable and $g^i=g_{m_in_i}(y_i) dy_i^{m_i}
\otimes dy_i^{n_i}$ is a metric on $M_i$ satisfying the equation
\ber
R_{m_in_i}[g^i]=\xi_{i}g^i_{m_{i}n_{i}},
\eer
$m_{i},n_{i}=1,\ldots,d_{i}$; $\xi_{i}=\const$, $i=0,\ldots,n$. The functions
$\gamma,\phi^{i}: {\bf R_\bullet}\rightarrow{\bf R}$ (${\bf R_\bullet}$
is an open subset of ${\bf R}$) are smooth.

We consider any manifold $M_i$ to be oriented and connected,
$i = 0,\ldots,n$. Then, the volume $d_i$-form
\beq
\label{2.9}
\tau_i=\sqrt{|g^i(y_i)|}\ dy_i^{1}\wedge\ldots\wedge dy_i^{d_i},
\eeq
and signature parameter $\eps(i)={\rm sign}({\rm det}(g^i_{m_{i}n_{i}}))=\pm1$
are correctly defined for all $i=0,\ldots,n$.

Let $\Omega$ from (\ref{2.1}) be a set of all non-empty subsets of
$\{0,\ldots,n\}$. The number of elements in $\Omega$ is
$|\Omega|=2^{n+1}-1$.

For any $I=\{i_1,\ldots,i_k\}\in\Omega$, $i_1<\ldots<i_k$, we put
in (\ref{2.2})
\beq
\label{2.10}
A^I=\Phi^I(u)\tau_{i_1}\wedge\ldots\wedge\tau_{i_k},
\eeq
where functions $\Phi^I:{\bf R_\bullet}\rightarrow{\bf R}$ are smooth,
$I\in\Omega$, and $\tau_{i}$ are defined in (\ref{2.9}). We denote by
\beq
d(I)\equiv d_{i_1}+\ldots+d_{i_k}=\sum_{i\in I}d_i
\eeq
the dimension of the oriented manifold $M_{I}=M_{i_1}\times\ldots
\times M_{i_k}$. It follows from (\ref{2.10}) that
\beq
\label{2.12}
F^I=dA^I=d\Phi^I\wedge\tau_{i_1}\wedge\ldots\wedge\tau_{i_k},
\mbox{ and } n_I=d(I)+1,
\eeq
$I \in \Omega$.

Thus, dimensions of forms $F^I$ in the considered model are fixed by a
subsequent decomposition of the manifold.

For dilatonic scalar fields we put $\varphi^I=\varphi^I(u)$, $I\in\Omega$.
Let
\beq
\label{2.13}
f=\gamma_0-\gamma, \quad \sum_{i=0}^{n}d_i\phi^i\equiv\gamma_0.
\eeq

It is not difficult to verify that the field equations
(\ref{2.4})--(\ref{2.5}) for field configurations from (\ref{2.8}),
(\ref{2.12}) may be obtained as equations of motion corresponding
to the action
\ber
\label{2.14}
S_{\sigma}=\frac{1}{2\kappa^{2}_0}
\int du\e{f}\biggl\{-wG_{ij}\dot\phi^i\dot\phi^j-
w\delta_{IJ}\dot\varphi^I\dot\varphi^J- \nn
w\sum_{I\in\Omega}\eps(I)\exp\Bigl(2\vec\lambda_I\vec\varphi-
2\sum_{i\in I}d_i\phi^i\Bigr)(\dot\Phi^I)^2-2V\e{-2f}\biggr\},
\eer
where $\vec\varphi=(\varphi^I)$, $\vec\lambda_I=(\lambda_{JI})$,
$\dot\varphi\equiv d\varphi(u)/du$; $G_{ij}=d_i\delta_{ij}-d_id_j$
are components of "pure cosmological" minisuperspace metric and
\beq
\label{2.15}
V={V}(\phi)=\Lambda\e{2{\gamma_0}(\phi)}-
\half\sum_{i =1}^{n}\xi_id_i\e{-2\phi^i+2 {\gamma_0}(\phi)}
\eeq
is the potential. In (\ref{2.14}) $\eps(I)\equiv\eps(i_1)\times\ldots\times
\eps(i_k)=\pm1$ for $I=\{i_1,\ldots,i_k\}\in\Omega$.

\section{Classical exact solutions}

We consider the harmonic time gauge $\gamma=\gamma_0(\phi)$. The problem
of integrability of the Lagrange equations arising in this model may be
simplified if we integrate the Maxwell equations
\ber
\label{3.2}
\frac d{du}\left(\exp\Bigl(2\vec\lambda_I\vec\varphi-
2\sum_{i\in I}d_i\phi^i\Bigr)
\dot\Phi^I\right)=0, \quad
\dot\Phi^I=Q^I\exp\left(-2\vec\lambda_I\vec\varphi+
2\sum_{i\in I}d_i\phi^i\right),
\eer
where $Q^I$ are constant, $I\in\Omega$.

Let $Q^I\ne0\Leftrightarrow I\in\Omega_*$, where $\Omega_*\subset\Omega$
is some non-empty subset of $\Omega$. For fixed $Q=(Q^I,I\in\Omega_*)$ the
Lagrange equations corresponding to $\phi^i$ and $\varphi^I$, when equations
(\ref{3.2}) are substituted, are equivalent to Lagrange equations for
the Lagrangian
\ber
\label{3.3}
L_Q=\frac12\left[G_{ij}\dot\phi^i\dot\phi^j+
\delta_{IJ}\dot\varphi^I\dot\varphi^J \right]-V_Q
\eer
where
\ber
\label{3.4}
V_Q=V_Q(\phi,\varphi,w)=-wV(\phi)+\frac12\sum_{I\in\Omega_*}
\eps(I)\exp\left(-2\vec\lambda_I\vec\varphi+
2\sum_{i\in I}d_i\phi^i\right)(Q^I)^2.
\eer
($V(\phi)$ is defined in (\ref{2.15})). Thus, we are led to the
pseudo-Euclidean Toda-like system (see \cite{IMZ}, \cite{IM3}--\cite{GIM})
with the zero-energy constraint:
\ber
\label{3.5}
L_Q=\frac12\bar G_{AB}\dot x^A\dot x^B-V_Q, \quad
E_Q=\frac12\bar G_{AB}\dot x^A\dot x^B+V_Q=0,
\eer
where $x=(x^A)=(\phi^i,\varphi^I)$,
\ber
\label{3.6}
(\bar G_{AB})=\left(\begin{array}{cc}
G_{ij}&0\\
0&\delta_{IJ}
\end{array}\right)
\eer
$i,j\in\{0,\dots,n\}$, $I,J\in\Omega$, and the potential $V_Q$ may be
presented in the following form
\ber
\label{3.7}
V_Q=\sum_{i=0}^n\left(\frac w2\xi_id_i\right)\exp[2U^i(x)]-
(w)\Lambda\exp[2U^\Lambda(x)] \nn
+\sum_{I\in\Omega_*}\frac{\eps(I)}2(Q^I)^2\exp[2U^I(x)],
\eer
where
\ber
\label{3.8}
U^\Lambda(x)=U_A^\Lambda x^A=\sum_{j=0}^nd_j\phi^j, \quad
U^i(x)=-\phi^i+\sum_{j=0}^nd_j\phi^j, \nn
U^I(x)=\sum_{i\in I}d_i\phi^i-\vec\lambda_I\vec\varphi,
\eer
$i,j\in\{0,\dots,n\}$, $I\in\Omega_*$, $J\in\Omega$. The contravariant
components of these vector read
\ber
U^{Ii}=G^{ij}U_j^I=\delta_i^I-\frac{d(I)}{D-2}, \quad
U^{\Lambda i}=\frac1{2-D},\quad U^{ji}=\frac{\delta^{ji}}{d_i},
\eer
where
\ber
\delta_i^I\equiv\sum_{j\in I}\delta_{ij}=\left\{\begin{array}{ll}
1,&i\in I\\
0,&i\notin I
\end{array}\right., \quad G^{ij}=\frac{\delta^{ij}}{d_i}+\frac1{2-D},
\eer
are the indicator of belonging $i$ to $I$ and the matrix inverse to the
matrix $(G_{ij})$ correspondingly, $i,j=0,\dots,n$.

Now let us continue with the case, when the scale factors of the
internal spaces are constant.

\section{The case of constant internal space factors}

The Lagrange system (\ref{3.5}) leads to the following equations of
motion
\ber
\label{4.1}
\bar G_{AB}\ddot x^B+\frac{\partial V_Q}{\partial x^A}=0.
\eer
Substituting the potential (\ref{3.7}) into (\ref{4.1}) we get the
set of equations
\ber
\label{4.3}
\ddot\phi^j-(w\xi_j)\exp[2U^j(x)]-
\frac{2w}{2-D}\Lambda\exp[2U^\Lambda(x)] \nn
+\sum_{I\in\Omega}\eps(I)(Q^I)^2\left(\delta_j^I-
\frac{d(I)}{D-2}\right)\exp[2U^I(x)]=0, \mm
\label{4.4}
\ddot\varphi^J-
\sum_{I\in\Omega}\eps(I)(Q^I)^2\lambda_I^J\exp[2U^I(x)]=0.
\eer
Now let us consider the special case with the $x^k=x_0^k=\const$,
$k=1,\dots,n$; $\varphi^I=\const$, $I\in\Omega$. This case corresponds
to the case with static internal spaces, when sizes of internal
dimensions may be set arbitrary small and so unobservable during the
whole evolution of the Universe. Then the equations of motion read
\ber
\label{4.5}
\ddot\phi^0-\xi_0A_0\exp[(2d_0-2)\phi^0]-
\Lambda A_\Lambda\exp[(2d_0)\phi^0]+
\sum_{I\in\Omega}A_I^0\exp[(2d_0\delta_0^I)\phi^0]=0, \mm
\label{4.6}
-\xi_kA_k\exp[(2d_0)\phi^0]-\Lambda A_\Lambda\exp[(2d_0)\phi^0]+
\sum_{I\in\Omega}A_I^k\exp[(2d_0\delta_0^I)\phi^0]=0, \mm
\label{4.6a}
\sum_{I\in\Omega}\lambda_I^JA_I^0\left(\delta_0^I-
\frac{d(I)}{D-2}\right)^{-1}\exp[(2d_0\delta_0^I)\phi^0]=0.
\eer
where $k=1,\dots,n$, $J\in\Omega$ and
\ber
\label{4.7}
A_0=w\prod_{l=1}^n(X^l)^{d_l}, \quad
A_k=\frac w{X^k}\prod_{l=1}^n(X^l)^{d_l}, \quad
A_\Lambda=\frac{2w}{2-D}\prod_{l=1}^n(X^l)^{d_l}, \mm
\label{4.8}
A_I^k=\eps(I)(\til Q^I)^2\left(\delta_k^I-\frac{d(I)}{D-2}\right)
\prod_{l\in I\setminus\{0\}}(X^l)^{d_l}.
\eer
where we introduce new variables $X^k$: $X^k\equiv\exp[2\phi^k]$,
$i=0,\dots,n$, $k=1,\dots,n$ and $\til Q^I=Q^I\exp[-\vec\lambda_I
\vec\varphi]$. Thus we have one differential equation (\ref{4.5}) for
$\phi^0$, the set of algebraic equations (\ref{4.6}) for $\phi^k$ and
equations (\ref{4.6a}) for $\varphi^I$, $I\in\Omega$.

As we can see, there are two possibilities to solve equations (\ref{4.5}),
(\ref{4.6}): $0\in I$, $\forall I\in\Omega$ or $0\notin I$,
$\forall I\in\Omega$, i.e. when forms are present in our external
space or they are completely defined on internal subspaces.

\subsection{The case with all $p$-branes "living" also in external space}

Let us consider the first case. The equation (\ref{4.5}) takes the form
\ber
\label{4.10}
\ddot\phi^0-\xi_0A_0\exp[(2d_0-2)\phi^0]+
\{-\Lambda A_\Lambda+A_\Omega^0\}\exp[(2d_0)\phi^0]=0, \quad
A_\Omega^0=\sum_{I\in\Omega}A_I^0.
\eer

The first integral of the equation (\ref{4.10}) reads
\ber
\label{4.11}
\frac{(\dot\phi^0)^2}2-\frac{\xi_0A_0}{2d_0-2}\exp[(2d_0-2)\phi^0]-
\frac{\Lambda A_\Lambda-A_\Omega^0}{2d_0}\exp[2d_0\phi^0]=E',
\eer
where due to the zero-energy constraint (\ref{3.5}) $E'$ takes the form
\ber
\label{4.11a}
-E'=\sum_{k=1}^n\frac{\xi_kd_k}{2d_0(1-d_0)}A_k\exp[(2d_0)\phi^0]+
\frac{\Lambda A_\Lambda}{2d_0}\left(\frac{D-2}{1-d_0}+1\right)
\exp[(2d_0)\phi^0] \nn
+\frac{\exp[(2d_0)\phi^0]}{2d_0}\sum_{I\in\Omega}A_I^0
\left[\frac1{1-d_0}\left(1-\frac{d(I)}{D-2}\right)^{-1}-1\right],
\eer
but if one sums the equations (\ref{4.6}) with the weights $d_k$, one can
get that $E'=0$. Thus, we have the zero-energy constraint for the subsystem
with $\phi^0$ that is usefull for the models with inflation. Finally we
get the following solution of (\ref{4.10})
\ber
\label{4.12}
\ints^{\phi^0}\frac{d\bar\phi^0}
{\sqrt{\ds\frac{\xi_0A_0}{d_0-1}\exp[(2d_0-2)\bar\phi^0]+
\ds\frac{\Lambda A_\Lambda-A_\Omega^0}{d_0}
\exp[2d_0\bar\phi^0]}}=u-u_0,
\eer
where $u_0=\const$.

The equation (\ref{4.6}) in this case takes the following form
\ber
\label{4.16}
\left(-\frac{w\xi_k}{X^k}-\frac{2\Lambda w}{2-D}\right)
\prod_{l=1}^n(X^l)^{d_l}+\sum_{I\in\Omega}\eps(I)(\til Q^I)^2
\left(\delta_k^I-\frac{d(I)}{D-2}\right)
\prod_{l\in I\setminus\{0\}}(X^l)^{d_l}=0,
\eer
$k=1,\dots,n$.

\subsection{The case with all $p$-branes not "living" in our space}

In the second case $0\notin I$, $\forall I\in\Omega$ to integrate
equations (\ref{4.6}) we must put the following restriction on the
scale factors of internal spaces
\ber
\label{4.17}
X^k=\xi_k\frac{D-2}{2\Lambda}.
\eer
This restriction let us cancel the terms with the scale factor of the $M_0$
space. Such a method called "fine-tuning" was previously used in \cite{BIMZ}
to obtain cosmological solutions with static internal spaces, but
without $p$-branes. So, we get simple equations for the internal space
factors:
\ber
\label{4.19}
\sum_{I\in\Omega}\eps(I)(\til Q^I)^2\left(\delta_k^I-\frac{d(I)}{D-2}\right)
\left(\frac{D-2}{2\Lambda}\right)^{d(I)}\prod_{l\in I}(\xi_l)^{d_l}=0.
\eer
$k=1,\dots,n$. Equation (\ref{4.5}) in this case takes the form
\ber
\label{4.20}
\ddot\phi^0-\xi_0A_0\exp[(2d_0-2)\phi^0]
-\Lambda A_\Lambda\exp[(2d_0)\phi^0]+A_\Omega^0=0.
\eer
But if we sum the equations (\ref{4.19}) with the weights $d_k$ we get that
$A_\Omega^0=0$ for this case. We also have the zero-energy constraint
for the subsystem with $\phi^0$ in this case. Thus the solution reads
\ber
\label{4.21}
\ints^{\phi^0}\frac{d\bar\phi^0}
{\sqrt{\ds\frac{\xi_0A_0}{d_0-1}\exp[(2d_0-2)\bar\phi^0]+
\ds\frac{\Lambda A_\Lambda}{d_0}\exp[2d_0\bar\phi^0]}}=u-u_0.
\eer
As we can see this solution does not depend on any $p$-brane properties.
So, this is the cosmological solution with $\Lambda$-term (see \cite{BIMZ}).
Thus, if $p$-branes do not "live" on $M_0$ and they are compensated in
internal spaces (see (\ref{4.16})), we can not "feel" the
influence of the branes configurations.

\section{Some examples and conclusion}

Now we consider some examples of the obtained solutions. As we can see,
only the first case when $0\in I$, $\forall I\in\Omega$ is interesting
as $p$-branes solutions. So, let us investigate the solutions in this
case. It is usefull to use the synchronous-time parametrization for the
solutions. Let $t_s$: $dt_s=e^\gamma du$, be a synchronous time. Here we put
$w=-1$. Then the solution (\ref{4.12}) takes the following form
\ber
\label{5.1}
\ints^{\phi^0}\frac{d\phi^0}{\sqrt{\ds\frac{\xi_0}{1-d_0}
\exp[-2\phi^0]-\ds\frac{\Lambda A_\Lambda-
A_\Omega^0}{d_0A_0}}}=t_s.
\eer

Let us consider some special case. Here we also use the "fine-tuning" of
a cosmological constant $\Lambda$, and put
\ber
\label{5.2}
\xi_0=d_0-1, \quad \xi_i=\frac{2\Lambda}{D-2},
\eer
$i=1,\dots,n$. Then from (\ref{5.1}) we get
\ber
\label{5.3}
g=-dt_s\otimes dt_s+\frac{\ch^2[Ht_s]}{H^2}g^0+\sum_{k=1}^ng^k,
\eer
where
\ber
\label{5.4}
d_0H^2=\frac{\Lambda A_\Lambda-A_\Omega^0}{-A_0}=
\frac{2\Lambda}{D-2}-\sum_{I\in\Omega}\eps(I)(\til Q^I)^2
\left(1-\frac{d(I)}{D-2}\right)\equiv\bar\xi_0.
\eer
Thus, for the new metric
\ber
\label{5.5}
\bar g^0\equiv-dt_s\otimes dt_s+\frac{\ch^2[Ht_s]}{H^2}g^0,
\eer
we have $\Ric[\bar g^0]=\bar\xi_0\bar g^0$. As we can see, for $(M_0,g^0)=
(S^{d_0},\ g[S^{d_0}]=d\Omega_{d_0}^2)$ we get the de Sitter space. In this case
the relations (\ref{4.6a}) and (\ref{4.16}) read
\ber
\label{5.6}
\sum_{I\in\Omega}\eps(I)(\til Q^I)^2\left(\delta_k^I-
\frac{d(I)}{D-2}\right)=0, \quad
\sum_{I\in\Omega}\lambda_I^J\eps(I)(\til Q^I)^2=0,
\eer
$k=1,\dots,n$, $J\in\Omega$. For the form $F^I$ we also get
\ber
\label{5.7}
F^I=\til Q^I\e{-\vec\lambda_I\vec\varphi}\bar\tau_0\wedge\tau(\bar I),
\eer
where $\bar\tau_0$ is the volume form on $dS^{d_0+1}$,
$\bar I=I\setminus\{0\}$, $I\in\Omega$.

{\bf Example.} Now let us consider some more special cases. Here we
put $n=2$ and
\ber
\label{5.8}
\Omega=\{\{0\},\{0,1\},\{0,2\}\}.
\eer
Thus, resolving the first equation in (\ref{5.6}) we get the following
formulas for the charges of forms
\ber
\label{5.9}
(\til Q^0)^2=-\frac{\eps(2)(d_0+1)}{d_0}\til Q^2, \quad
(\til Q^{0,1})^2=\frac{\eps(2)}{\eps(1)}\til Q^2, \quad
(\til Q^{0,2})^2=\til Q^2,
\eer
where $\til Q^2$ is some arbitrary constant. Here, as we can see,
one must put $\eps(1)=\eps(2)=-1$. Thus, the effective cosmological
term of the $\bar g^0$ space in this case reads
\ber
\label{5.10}
\frac{2\til\Lambda}{d_0-1}=\frac{2\Lambda}{D-2}-\frac{\eps(0)}{d_0}\til Q^2,
\eer
so, for the case with $(M_0,g^0)=(S^{d_0},\ g[S^{d_0}]=d\Omega_{d_0}^2)$
($\bar g^0$ is de Sitter space) $\eps(0)=+1$ and we may get the extremely
small effective cosmological constant $\til\Lambda$ while the multidimensional
cosmological bare constant $\Lambda$ has the Planck scale.

{\bf The case with $\Lambda=0$.} If we put $\Lambda=0$, then due to
the equation (\ref{4.16}) we come to the case with Ricci-flat internal
spaces, i.e. $\xi_k=0$, $k=1,\dots,n$, without any other changes in
the results of the section 5. But inflationary solutions are
generated in this case only due to the fields of forms ($p$-branes).

Thus, we considered the multidimensional cosmology, with $n+1$ Einstein
spaces of non-zero curvature $(M_i,g^i)$, $i=0,\dots,n$ in the presence
of several scalar fields and forms. When scale factors of the internal
spaces $(M_k,g^k)$, $k=1,\dots,n$, are chosen to be constant, we obtained
the exact solution, describing the evolution of one external space
$(M_0,g^0)$ in the presence of $n$ "frozen" internal spaces. There were
investigated two possible cases: $0\in I$, $\forall I\in\Omega$ and
$0\notin I$, $\forall I\in\Omega$. The second case turned out to be off
interest because  $p$-brane configurations do not have any influence
on a scale factor of the external space. In this case we had an ordinary
cosmological solution with $\Lambda$-term for $M_0$.

In the other case the influence of $p$-branes is rather important.
There we got the solution for the metric of $M_0$ in the synchronous-time
parametrization, that allowed us to investigate this solution in detail.
Although we considered the de Sitter type solution, we may also obtain
the anti-de Sitter one for another sign of curvature. So, the solution
may be de Sitter or anti-de Sitter one depending on $p$-brane configurations.
In this case it is also necessary to use "fine-tuning" of the cosmological
constant and curvatures of internal spaces to obtain the solutions. This
allows us to make the effective cosmological constant $\til\Lambda$ extremely
small while the multidimensional cosmological constant $\Lambda$ has the
Planck scale.

\begin{center}
{\bf Acknowledgments}
\end{center}

This work was supported in part by the DFG grant 436 RUS 113/236/O(R),
by the Russian Ministry of Science and Technology and by RFBR grant
98-02-16414.

\end{document}